\documentclass[fleqn,twoside]{article}
\usepackage{espcrc2}

\usepackage{graphicx}
\usepackage[figuresright]{rotating}

\newcommand{\AmS}{{\protect\the\textfont2
  A\kern-.1667em\lower.5ex\hbox{M}\kern-.125emS}}

\newcommand{\bea}{\begin{eqnarray}}
\newcommand{\eea}{\end{eqnarray}}

\title{Mass and Spin Measurement with $M_{T2}$ and MAOS Momentum}

\author{Won Sang Cho\address{Department of Physics and Astronomy, Seoul National
University,\\
        Seoul 151-747, Korea},
        Kiwoon Choi\address[MCSD]{Department of Physics, Korea Advanced Institute of Science and Technology,\\ Daejeon 305-701, Korea},
        Yeong Gyun Kim\address{Physics Division, Korea Institute for Advanced
        Study, \\
        Seoul 130-722, Korea}
        and
        Chan Beom Park\addressmark[MCSD]\thanks{
          This work is supported by the NRF
grants funded by the Korean Government (KRF-2007-341-C00010,
KRF-2008-314-C00064).}}

\begin{document}

\begin{abstract}
We discuss the $M_{T2}$-kink method to determine the masses of both
the dark matter WIMP and its mother particle produced at the LHC. We
then introduce a new kinematic variable, the
$M_{T2}$-Assisted-On-Shell (MAOS) momentum, that provides a
systematic approximation to the invisible particle momenta in hadron
collider events producing a pair of invisible particles, and apply
it to certain SUSY processes and their UED equivalents to determine
the spin of gluino/KK-gluon and of slepton/KK-lepton. An application
of the MAOS momentum to the SM Higgs mass measurement is briefly
discussed also. \vspace{1pc}
\end{abstract}

\maketitle

\section{INTRODUCTION}

The Large Hadron Collider (LHC) at CERN will explore soon the TeV
energy scale where new physics beyond the Standard Model (SM) is
likely to reveal itself. There are two major motivations for new
physics at the TeV scale, one is the hierarchy problem and the other
is the existence of dark matter (DM). Constraints from electroweak
precision measurements and proton decay suggest that TeV scale new
physics preserves a $Z_2$ parity under which new particles are odd,
while the SM particles are even. Well known examples include the
weak scale supersymmetry (SUSY) with conserved $R$-parity
\cite{susy}, little Higgs model with $T$-parity \cite{littlehiggs},
and universal extra dimension (UED) model with $KK$-parity
\cite{ued}. The lightest new particle in these $Z_2$-preserving
models is typically a weakly interacting massive particle (WIMP)
which is a good DM candidate. These new physics models also predict
a clear LHC signature: significant excess of multi-jet (possibly
with isolated leptons) events with large missing transverse
momentum, which would be an indication of pair-produced new
particles decaying to visible SM particles and a pair of invisible
WIMPs (see Fig.~1).

To clarify the underlying theory of new physics events, it is
crucial to measure the mass and spin of new particles produced at
the LHC. However it is quite challenging to measure the unknown mass
in such missing energy events as (i) the initial parton momenta in
the beam direction are unknown in hadron collider experiments and
(ii) each event involves two invisible WIMPs in the final state.
Spin measurement is likely to be even more challenging as it
typically requires a more refined event reconstruction and/or a
polarized mother particle state. In this talk, we wish to discuss
the recently proposed $M_{T2}$-kink method of mass measurement
\cite{mt2kink1,mt2kink2}, which can be applied in principle to a
wide class of new physics processes with the event topology of
Fig.~1.
 We also introduce a new collider variable, the
$M_{T2}$-Assisted-On-Shell (MAOS) momentum \cite{maos}, providing a
systematic approximation to the invisible momenta in the events of
Fig.~1, and apply it to certain SUSY and UED processes to determine
the spin of gluino/KK-gluon and of slepton/KK-lepton. Finally we
briefly discuss the application of the MAOS momentum to the SM Higgs
mass measurement \cite{maoshiggs}.

\begin{figure}[ht!]
\begin{center}
\includegraphics[width=15pc]{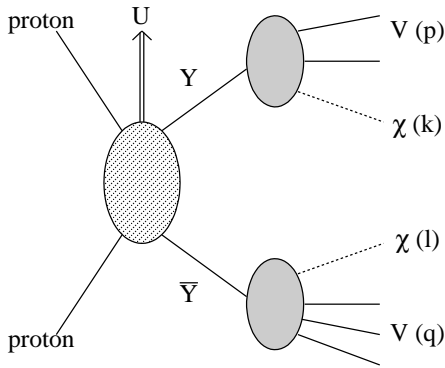}
\end{center}
\vskip -1.3cm \caption{New physics events at the LHC producing a
pair of invisible WIMPs.} \label{event_topology}
\end{figure}

\section{$M_{T2}$ KINK FOR MASS MEASUREMENT}

There are several kinematic methods proposed so far to determine the
unknown  masses in hadron collider experiments producing a pair of
invisible particles.  A simple and perhaps the most well-known
method is the endpoint method \cite{endpoint} which relies on that
the endpoint value of the invariant mass distribution of visible
(SM) decay products in each decay chain depend on the involved new
particle masses. For example, if one considers a cascade decay
$Y\rightarrow V_1+I\rightarrow V_1+V_2+\chi$, where $I$ is an
intermediate on-shell particle, $V_1$ and $V_2$ are massless visible
particles, and $\chi$ is an invisible particle, the endpoint of the
invariant mass of $V_1+V_2$ is given by \bea m_{V_1V_2}^{\rm
max}=m_{Y}\sqrt{\left(1-m_{I}^2/m_{Y}^2\right)\left(1-m_{\chi}^2/m_{I}^2\right)}.
\eea On the other hand, for a single step 3-body decay $Y\rightarrow
V_1+V_2+\chi$ without an intermediate on-shell particle, one finds
\bea m_{V_1V_2}^{\rm max}=m_Y-m_\chi. \eea If one considers a
$n$-step decay chain, $$Y\rightarrow V_1+I_1\rightarrow
...\rightarrow V_1+...+V_n +\chi,$$ there are $(n+1)$ unknown
masses, i.e. $m_Y,m_\chi, m_{I_n}$ $(n=1,2,...,n-1)$, and
$2^n-(n+1)$ invariant mass combinations of visible particles. Then
for the case with $n\geq 3$, there can be enough number of endpoint
values with which the unknown masses are completely determined.

An alternative method that might be applied to  a long decay chain
is the mass relation method which attempts to reconstruct the
missing momenta using the on-shell constraints imposed on multiple
events with the same event topology \cite{mass_relation}. For
$n$-step cascade decays, each event has $n+1$ unknown masses and one
invisible WIMP momentum, and is constrained by  $n+1$ on-shell
conditions. For multiple ($N>1$) events, the unknown masses are
common, so the total number of unknowns is $n+1+4N$. On the other
hand, the total number of on-shell constraints increases as
$N(n+1)$. This implies that the WIMP momentum of each event might be
reconstructed  (but generically with a discrete degeneracy) if
$n\geq 4$ and $N\geq (n+1)/(n-3)$. If the events are made of a
symmetric pair of $n$-step cascade decays and the missing transverse
momentum ${\bf p}_T^{\rm miss}$ of each event is available, the WIMP
momentum can be reconstructed (with discrete degeneracy)  even when
$n=3$.

As noticed above, the mass relation method requires a long decay
chain  with $n\geq 3$.
%  to have enough number of independent
%on-shell constraints from which one may reconstruct the missing
%momenta.
Such a long decay chain is required for the endpoint method also if
one wishes to determine the absolute mass scale of new physics.
%enough number of independent invariant mass distributions whose
%endpoint values can completely determine the involved new particle
%masses.
On the other hand, typical new physics models contain a rather large
parameter region which does not give a long decay chain with $n\geq
3$.
 For instance, in $R$-parity conserving SUSY with
the mSUGRA boundary condition defined by the universal sfermion mass
$m_0$  and the universal gaugino mass $M_{1/2}$ at the GUT scale,
the sleptons $\tilde{\ell}$ will be heavier than the second lightest
neutralino $\chi_2$ if $m_0> 0.8 M_{1/2}$. In this case, the popular
3-step squark decay, $\tilde{q}\rightarrow q{\chi_2}\rightarrow
q\ell \tilde{\ell}^*\rightarrow  q\ell\bar{\ell}\chi_1$, is not
available anymore. Also some string-motivated SUSY breaking schemes
predict that squarks and sleptons have masses heavier than few TeVs,
while gauginos have light masses in sub-TeV range
\cite{gaugino_code}. In such case, only gluinos will be copiously
produced at the LHC, and then  followed by the single step 3-body
decay $\tilde{g}\rightarrow q\bar{q}\chi_1$ or the 2-step 3-body
decays $\tilde{g}\rightarrow q\bar{q}\chi_2\rightarrow
q\bar{q}\ell\bar{\ell}\chi_1$. In this case, the mass relation
method can not be applied  at all, and the endpoint method
determines only the mass differences $m_{\tilde{g}}-m_{\chi_1}$ and
$m_{\chi_2}-m_{\chi_1}$.  On the other hand, the $M_{T2}$-kink
method which will be discussed below can determine the full gaugino
mass spectrum including the absolute mass scale.

The $M_{T2}$ kink method  is based on the observation that the
endpoint of the collider variable $M_{T2}$ \cite{lester}, when
considered as a function of the trial WIMP mass $\tilde{m}_\chi$,
generically exhibits a kink structure at $\tilde{m}_\chi=m_\chi^{\rm
true}$ \cite{mt2kink1,mt2kink2}. Unlike the endpoint method and the
mass relation method, it can in principle determine the unknown
masses
 even when the mother particle experiences only single-step decay or two-step
cascade decays.

The collider variable $M_{T2}$ is a generalization of the transverse
mass  to the event with two chains of decays each of which produces
an  invisible particles as Fig.~1. So let us first consider the
transverse mass variable $M_T$ defined for a decay process $$
Y\rightarrow V(p)+\chi(k),$$ where $V(p)$ denotes a set of
(generically arbitrary number of) visible particles with the total
momentum $p$ and $\chi(k)$ is the invisible WIMP with momentum $k$.
This process can be either a single step decay or a multi-step decay
chain. The transverse mass of the mother particle $Y$ is defined as
\bea M_T^2(Y)=p^2+\tilde{m}_\chi^2+2E_T(p)E_T(k)-2{\bf p}_T\cdot
{\bf k}_T, \eea where $\tilde{m}_\chi$ is the trial WIMP mass, ${\bf
p}_T$ and ${\bf k}_T$ are the transverse momenta, and
$E_T(p)=\sqrt{p^2+|{\bf p}_T|^2}$ and
$E_T(k)=\sqrt{\tilde{m}_\chi^2+|{\bf k}_T|^2}$ are the transverse
energies of $V$ and $\chi$, respectively. Note that $M_T$ is a
monotonically increasing function of $\tilde{m}_\chi$, and in the
zero-width approximation its value at $\tilde{m}_\chi=m_\chi$ is
bounded as
\begin{eqnarray}
M_T^2(Y;{\tilde{m}_\chi=m_\chi})\, \leq\,  m_Y^2, \nonumber
\end{eqnarray}
where $m_\chi$ is the true WIMP mass, and $m_Y^2$ denotes the
invariant mass peak of $(p+k)^2=\,p^2+m_\chi^2+2E_T(p)E_T(k)\cosh
(\eta_p-\eta_k)-2{\bf p}_T\cdot {\bf k}_T$, where
$\eta=\frac{1}{2}\ln(E+p_z)/(E-p_z)$.

Let us now consider a new physics event \bea \label{new_physics} &&
Y(p+k)
+\bar{Y}(q+l)+U(u)\nonumber \\
&\rightarrow& V(p)+\chi(k)+V(q)+\chi(l)+U(u),\nonumber \eea where
$U(u)$ stands for visible particles not associated with the decays
of $Y$ and $\bar{Y}$, which carry the total momentum $u$ that will
be called the upstream momentum (see Fig.~1). The event variable
$M_{T2}$ is defined as
\begin{equation}
M_{T2}= \min_{\mathbf{k}_{T}+\mathbf{l}_{T}=\mathbf{p}_T^{\rm miss}}
\Big[ \mathrm{max}\Big\{M_{T}(Y), M_{T}(\bar{Y})\Big\}\Big],
\label{mt2_def}
\end{equation}
where the missing transverse momentum is given by $\mathbf{p}_T^{\rm
miss}=-(\bf{p}_T+\bf{q}_T+\bf{u}_T)$. Then $M_{T2}$ can be
considered as a function of the measurable event variables $p^2,
{\bf p}_T, q^2, {\bf q}_T, {\bf u}_T$, and also of the trial WIMP
mass $\tilde{m}_\chi$. Obviously, by construction $M_{T2}$ is a
monotonically increasing function of the trial WIMP mass
$\tilde{m}_\chi$, and is bounded as \bea
M_{T2}(\tilde{m}_\chi=m_\chi)\leq  m_Y,\eea in the zero-width
approximation. Generically there can be multiple events whose
$M_{T2}$  saturate the upper bound at $\tilde{m}_\chi=m_\chi$, i.e.
$M_{T2}(\tilde{m}_\chi=m_\chi)=m_Y$, but the corresponding
$M_{T2}(\tilde{m}_\chi)$ have different slopes at
$\tilde{m}_\chi=m_\chi$. This simple observation suggests that the
endpoint values of $M_{T2}$,
%\bea M_{T2}^{\rm
%max}(\tilde{m_\chi})= {\rm max}_{\mbox{all events}}\Big[
%M_{T2}(p,q;\tilde{m}_\chi)\Big] \eea
\[ M_{T2}^{\rm max}(\tilde{m}_\chi)=\max_{p,q,u}
\Big[\,M_{T2}(p^2,{\bf p}_T,q^2,{\bf q}_T,{\bf
u}_T;\tilde{m}_\chi)\,\Big],
\] generically exhibits  a kink  at $\tilde{m}_\chi=m_\chi$.

The kink structure of $M_{T2}^{\rm max}$ at $\tilde{m}_\chi=m_\chi$
can be understood more clearly with an explicit analytic expression
of $M_{T2}$. Although not available for generic events,  the
analytic expression of $M_{T2}$ can be easily obtained for a special
type of events including the events with vanishing upstream
transverse momentum, ${\bf u}_T=0$, and also the symmetric events
with $p^2=q^2$, ${\bf p}_T={\bf q}_T$ and arbitrary ${\bf u}_T$. The
$M_{T2}$-kink has been studied in \cite{mt2kink1} using the analytic
expression of $M_{T2}$ for the event set with ${\bf u}_T=0$, which
has been derived in \cite{lester,mt2kink1}. Here we will consider a
different event set consists of symmetric events with arbitrary
${\bf u}_T$, which has a simple analytic expression of $M_{T2}$
explaining  the origin and structure of kink.

For symmetric events with $$p^2=q^2, \quad {\bf p}_T={\bf q}_T,$$
one easily finds \bea M_{T2}=M_T(p^2,{\bf p}_T, \tilde{m}_\chi, {\bf
k}_T=-{\bf p}_T-\frac{{\bf u}_T}{2}) \eea and thus \bea &&\hskip
-0.7cm M_{T2}^2=-\,\frac{|{\bf
u}_T|^2}{4}\nonumber \\
&& \hskip -0.65cm +\left[\sqrt{p^2+|{\bf p}_T|^2}+
\sqrt{\tilde{m}_\chi^2+\left|{\bf p}_T+\frac{{\bf
u}_T}{2}\right|^2}\right]^{2}. \eea We are interested in the
endpoint events at $\tilde{m}_\chi=m_\chi$, i.e. the events with
$$ M_{T2}(\tilde{m}_\chi=m_\chi)=m_Y.$$ It is then straightforward
to find that $M_{T2}$ of such endpoint events can be parameterized
by two event variables $E_T\equiv \sqrt{p^2+|{\bf p}_T|^2}$ and
$u\equiv |{\bf u}_T|$: \bea &&\hskip -0.70cm
M_{T2}^2=-\,\frac{u^2}{4}\nonumber
\\
&& \hskip -0.65cm
+\left[E_T+\sqrt{\left(\sqrt{m_Y^2+\frac{u^2}{4}}-E_T\right)^2+\tilde{m}_\chi^2-m_\chi^2}\right]^2,\nonumber
\eea from which one obtains \bea
\left.\frac{dM_{T2}}{d\tilde{m}_\chi}\right|_{\tilde{m}_\chi=m_\chi}=\,
\left(\frac{m_\chi}{m_Y}\right)\left(\frac{1}{1-{\cal
E}_T}\right),\label{slope} \eea where $$ {\cal E}_T\equiv
\frac{E_T}{\sqrt{m_Y^2+u^2/4}}.$$ This shows that if some endpoint
events at $\tilde{m}_\chi=m_\chi$ have different values of the event
variable ${\cal E}_T$, their $M_{T2}$ curves have  different slopes
at $\tilde{m}_\chi=m_\chi$, so exhibit a kink.

Eq.~(\ref{slope}) indicates that the shape of the $M_{T2}$-kink is
determined by the range of ${\cal E}_T=E_T/\sqrt{m_Y^2+u^2/4}$
covered by the endpoint events at $\tilde{m}_\chi=m_\chi$.  From the
endpoint condition $M_{T2}(\tilde{m}_\chi=m_\chi)=m_Y$,  one can
find that the transverse energy $E_T$  is bounded (for given values
of $u$ and $p^2$) as \bea \hskip 2cm E_T^-\leq E_T \leq
E_T^+,\label{epm}\nonumber \eea where \bea &&\hskip -0.65cm
E_T^{\pm}=\frac{1}{2m_Y^2}\left[(m_Y^2-m_\chi^2+p^2)
\sqrt{m_Y^2+\frac{u^2}{4}} \right.\nonumber \\
&&\hskip
-0.65cm\left.\pm\frac{u}{2}\sqrt{(m_Y-m_\chi)^2-p^2}\sqrt{(m_Y+m_\chi)^2-p^2}\right].
\nonumber\eea Combining this with that the visible invariant mass
$\sqrt{p^2}$ in the decay process $Y\rightarrow V(p)+\chi$ is
bounded as
$$
0\leq \sqrt{p^2} \leq m_Y-m_\chi,
$$
we find that the possible range of ${\cal E}_T$
is given by \bea \hskip 2cm {\cal E}_T^{\rm min}\leq {\cal E}_T \leq
{\cal E}_T^{\rm max},\eea where
 \bea
{\cal E}_T^{\rm min}
%&\hskip -0.2cm=&\hskip -0.2cmE_T(p^2=0) \nonumber\\
&\hskip -0.2cm=&\hskip -0.2cm \left(
1-\frac{u}{2\sqrt{m_Y^2+u^2/4}}\right)\left(\frac{m_Y^2-m_\chi^2}{2m_Y^2}\right),
\nonumber \\
{\cal E}_T^{\rm max}
%&\hskip -0.2cm=&\hskip
%-0.2cmE_T(p^2=m_Y^2-2m_\chi\sqrt{m_Y^2+\frac{u^2}{4}}+m_\chi^2)\nonumber
%\\
&\hskip -0.2cm=&\hskip -0.2cm 1-\frac{m_\chi}{\sqrt{m_Y^2+u^2/4}}
\nonumber\eea for
$$
0\leq u \leq \frac{m_Y^2-m_\chi^2}{m_\chi},
$$
and \bea {\cal E}_T^{\rm min}
%&\hskip -0.2cm=&\hskip -0.2cmE_T^+(p^2=0)
%\nonumber
%\\
&\hskip -0.2cm=&\hskip
-0.2cm\left(1-\frac{u}{2\sqrt{m_Y^2+u^2/4}}\right)\left(\frac{m_Y^2-m_\chi^2}{2m_Y^2}\right),
\nonumber \\ {\cal E}_T^{\rm max}
%&=&E_T^-(p^2=0)\nonumber
%\\
&\hskip -0.2cm=&\hskip
-0.2cm\left(1+\frac{u}{2\sqrt{m_Y^2+u^2/4}}\right)\left(\frac{m_Y^2-m_\chi^2}{2m_Y^2}\right)\nonumber
\eea for
$$
u\geq \frac{m_Y^2-m_\chi^2}{m_\chi}.
$$

Note that ${\cal E}_T^{\rm max}$ and ${\cal E}_T^{\rm min}$
correspond to the upper and lower bounds of ${\cal E}_T$ for generic
symmetric endpoint events having a fixed value of $u$. The actual
range of ${\cal E}_T$ for a specific set of events might be
significantly narrower than the range defined by ${\cal E}_T^{\rm
max}$ and ${\cal E}_T^{\rm min}$.
%a wide class of new physics events expected at the LHC. However
Still the above discussion implies  that the appearance of kink is
quite generic, although its shape can differ for different sets of
new physics events. It implies also that an important factor to
determine the shape of kink is the range of $p^2$ covered by the
event set, which depends on whether $V(p)$ is a single particle
state or a multi-particle state and also on whether $Y\rightarrow
V+\chi$ is a single step decay or a multi-step decay chain
\cite{mt2kink1}. The upstream momentum also affects the shape of
kink significantly if $u$ is as large as ${\cal O}(m_Y)$
\cite{mt2kink2}.

\begin{figure}[ht!]
\begin{center}
\includegraphics[width=13pc]{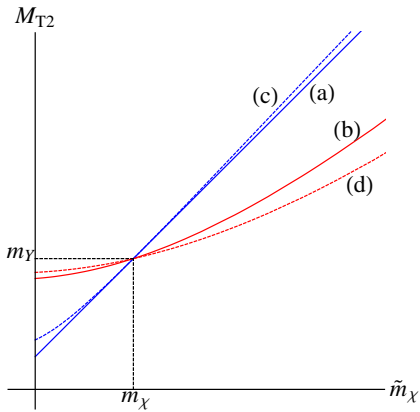}
\end{center}
\vskip -1.3cm \caption{$M_{T2}(\tilde{m}_\chi)$-curves for some
endpoint events at $\tilde{m}_\chi=m_\chi$, which explain the origin
of kink.} \label{kink_origin}
\end{figure}

In Fig.~2, we depict $M_{T2}(\tilde{m}_\chi)$ for the new physics
process:
$$Y+\bar{Y}+U\rightarrow V_1V_2\chi+ V_1V_2\chi+ U,$$
where $V_i$ $(i=1,2)$ are massless visible particles with 4-momentum
$p_i$, $U$ denotes a generic upstream momentum, and
$p^2=(p_1+p_2)^2$ can have any value between 0 and $(m_Y-m_\chi)^2$.
The 4 curves in Fig.~2 represent $M_{T2}(\tilde{m}_\chi)$ of the
following 4 endpoint events at $\tilde{m}_\chi=m_\chi$: (a) $u=0,\,
p^2=(m_Y-m_\chi)^2$, (b) $u=0,\, p^2=0$, (c) $u=m_Y,\,
p^2=(m_Y-m_\chi)^2, \cos\theta=-1$ (d) $u=m_Y,\, p^2=0,
\cos\theta=1$, where $\theta$ is the angle between ${\bf p}_T$ and
${\bf u}_T$, and we choose $m_Y/m_\chi=6$.

Of course, to make $M_{T2}$ a viable observable for real collider
data, one needs to isolate the new physics events from backgrounds
and also resolve the associated combinatoric ambiguities
\cite{mt2application}. In some cases, the $M_{T2}$ kink determined
by the curves (a) and (b) in Fig.~2 can be reproduced well even in
realistic Monte Carlo analysis including the detector effects and
the combinatoric errors \cite{mt2kink1,nojiri}. In Fig.~3, we
present $M_{T2}^{\rm max}(\tilde{m}_\chi)$ obtained in
\cite{mt2kink1} for the process $\tilde{g}+\tilde{g}\rightarrow
q\bar{q}\chi +q\bar{q}\chi$ under the assumption that the gaugino
masses take the anomaly pattern \cite{gaugino_code} with
$m_{\tilde{g}}=780$ GeV and all sfermions have a mass around few
TeV. It is in principle possible to apply the $M_{T2}$-kink method
to a wide class of new physics processes with the event topology of
Fig.~1. However it requires a detailed case-by-case study to see
whether or not the $M_{T2}$-kink method can be successfully
implemented in each case.

\begin{figure}[ht!]
\begin{center}
\includegraphics[width=15pc]{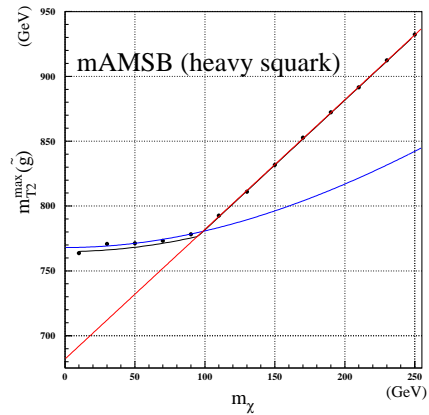}
\end{center}
\vskip -1.3cm \caption{$M_{T2}$-kink for the anomaly pattern of
gaugino masses in heavy sfermion scenario, including the effects of
combinatoric errors and detector smearing.} \label{kink_anomaly}
\end{figure}

\section{MAOS MOMENTUM AND SPIN MEASUREMENT}

The $M_{T2}$-Assisted-On-Shell (MAOS) momentum is an event variable
that approximates systematically the invisible momenta in hadron
collider events producing a pair of invisible particles as Fig.~1
\cite{maos,maoshiggs}:
$$ Y+\bar{Y}+U\rightarrow V(p)\chi(k)+V(q)\chi(l)+U(u).$$ For each
event of this type, the transverse MAOS momenta, ${\bf k}_T^{\rm
maos}$ and ${\bf l}_T^{\rm maos}$, correspond to the transverse
components of the trial WIMP momenta which determine $M_{T2}$, i.e.
the solution of  \bea && \hskip -0.65cm M_T(q^2,{\bf
q}_T,\tilde{m}_\chi, {\bf l}_T^{\rm maos})\leq
M_T(p^2,{\bf p}_T,\tilde{m}_\chi, {\bf k}_T^{\rm maos})\nonumber \\
&&\hskip -0.65cm =M_{T2}(p^2,{\bf p}_T,q^2,{\bf q}_T,{\bf
u}_T;\tilde{m}_\chi)\label{tmaos}\eea under the constraint \bea {\bf
k}^{\rm maos}_T+{\bf l}^{\rm maos}_T=-{\bf p}_T-{\bf q}_T-{\bf
u}_T\equiv {\bf p}_T^{\rm miss}. \label{missingpt} \eea Note that
here the first inequality is just a convention to define $p$ and
$q$.

As for the longitudinal and energy components, one can consider two
schemes. The first is to use the solution of the on-shell conditions
for both $\chi$ and $Y$:
 \bea
\label{maos1} && \hskip -0.65cm k^2_{\rm maos}=l^2_{\rm
maos}=m_\chi^2,\nonumber
\\
&&\hskip -0.65cm (p+k_{\rm maos})^2=(q+l_{\rm maos})^2=m_Y^2,
  \eea
and the second uses the solution of \bea \label{maos2} && \hskip
-0.65cm k^2_{\rm maos}=l^2_{\rm maos}=m_\chi^2,\nonumber
\\
&&\hskip -0.65cm \frac{k^{\rm maos}_z}{k_0^{\rm
maos}}=\frac{p_z}{p_0}, \quad \frac{l^{\rm maos}_z}{l_0^{\rm
maos}}=\frac{q_z}{q_0}. \eea  If both mother particles, $Y$ and
$\bar{Y}$, are in on-shell, the MAOS momenta obtained in both
schemes provide a useful approximation to the true WIMP momenta. On
the other hand, if any of $Y$ and $\bar{Y}$ is in off-shell, only
the second scheme provides an acceptable approximation.

In case that $m_\chi$ and $m_Y$ are unknown, one can use $m_\chi=0$
and $m_Y=M_{T2}^{\rm max}(m_\chi=0)$ instead of the unknown true
masses, and the resulting MAOS momenta still might be a reasonable
approximation to the true WIMP momenta  if $m_\chi^2/m_Y^2\ll 1$.
 Here we simply assume that $m_\chi$ and $m_Y$ are known, and focus on
 the first scheme defined by Eqs. (\ref{tmaos}), (\ref{missingpt}) and (\ref{maos1}).
In the next section, we will briefly discuss the application of the
second scheme to the Higgs mass measurement with the process
$H\rightarrow WW\rightarrow \ell\nu\bar{\ell}\nu$.

The transverse MAOS momenta are uniquely determined by (\ref{tmaos})
and (\ref{missingpt}). On the other hand,  each of the longitudinal
MAOS momenta determined by (\ref{maos1}) generically has two-fold
degeneracy: \bea &&\hskip -0.65cm k_z^{\rm maos}(\pm)=\frac{Ap_z\pm
p_0\sqrt{A^2-E_T^2(p)E_T^2(k)}}{p^2+|{\bf p}_T|^2}, \nonumber
\\ &&\hskip -0.65cm l_z^{\rm maos}(\pm)=\frac{Bq_z\pm
q_0\sqrt{B^2-E_T^2(q)E_T^2(l)}}{q^2+|{\bf q}_T|^2}, \eea where \bea
&&\hskip -0.65cm A=\frac{m_Y^2-m_\chi^2-p^2}{2}+{\bf p}_T\cdot {\bf
k}_T^{\rm maos}, \nonumber \\
&&\hskip -0.65cm B=\frac{m_Y^2-m_\chi^2-q^2}{2}+{\bf q}_T\cdot {\bf
l}_T^{\rm
maos}, \nonumber \\
&&\hskip -0.65cm E^2_T(p)=p^2+|{\bf p}_T|^2,\,\, E^2_T(q)=q^2+|{\bf q}_T|^2, \nonumber \\
&&\hskip -0.65cm E^2_T(k)=m_\chi^2+|{\bf k}_T^{\rm maos}|^2, \,\,
E^2_T(l)=m_\chi^2+|{\bf l}_T^{\rm maos}|^2.\nonumber \eea
%&&B=\frac{m_Y^2-m_\chi^2-q^2}{2}+{\bf q}_T\cdot {\bf l}_T^{\rm maos}
%\nonumber \eea and
%Note that $l_z^{\rm maos}$ can be obtained by replacing $p^\mu,{\bf
%k}_T^{\rm maos},A$ with $q^\mu,{\bf l}_T^{\rm maos}$.
Obviously both
${k}_z^{\rm maos}$ and ${l}_z^{\rm maos}$ are real
 iff
 \bea |A|\geq E_T(p)E_T(k), \quad |B|\geq E_T(q)E_T(l),\nonumber
 \eea
  which is equivalent to \bea
\label{realcondition} m_Y \geq \mathrm{max}\Big\{M_{T}(Y),
M_{T}(\bar{Y})\Big\},\eea
 where $M_T(Y)$ and $M_T(\bar{Y})$ are the transverse masses of $Y\rightarrow V(p)+\chi(k)$
 and $\bar{Y}\rightarrow V(q)+\chi(l)$, respectively, for ${\bf
 k}_T={\bf k}_T^{\rm maos}$ and ${\bf l}_T={\bf l}_T^{\rm maos}$.
This condition is always satisfied when the correct values of $m_Y$
and $m_\chi$ are used, and thus the MAOS momenta are real for all
events once constructed using the true WIMP and mother particle
masses.

For the endpoint events of balanced $M_{T2}$ \cite{lester,mt2kink1},
one has $M_T(Y)=M_T(\bar{Y})=m_Y$ and thus $|A|=E_T(p)E_T(k)$ and
$|B|= E_T(q)E_T(l)$. Obviously then both $k_\mu^{\rm maos}$ and
$l_\mu^{\rm maos}$ correspond to the unique solution of Eqs.
(\ref{tmaos}), (\ref{missingpt}) and (\ref{maos1}). In this case,
the true WIMP momenta also satisfy the same equations, which means
\bea k^{\rm maos}_\mu = k^{\rm true}_\mu, \quad l^{\rm
maos}_\mu=l^{\rm true}_\mu \eea for the endpoint events of balanced
$M_{T2}$. On the other hand, the endpoint events of unbalanced
$M_{T2}$ have  $M_T(\bar{Y})<M_T(Y)=m_Y$, so only $k_\mu^{\rm maos}$
corresponds to the true WIMP momentum.
 This observation suggests
that the MAOS momenta may approximate well the true WIMP momenta at
least for an appropriate subset of events near the $M_{T2}$ endpoint
\cite{maos}.

With a Monte Carlo analysis, one can confirm  that this is indeed
true.  As an example, we have examined the distribution of
$$\frac{\Delta {\bf k}_T}{{\bf k}^{\rm true}_T} \equiv \frac{\tilde{\bf k}_T-{\bf k}^{\rm true}_T}{{\bf
k}_T^{\rm true}}$$ for the gluino pair decay process:
$\tilde{g}+\tilde{g}\rightarrow q\bar{q}\chi+ q\bar{q}\chi$. Fig.~4
shows the results obtained for the focus (SPS2) point of mSUGRA
scenario: the dotted line is the distribution of $\Delta {\bf
k}_T/{\bf k}^{\rm true}_T$ for $\tilde{\bf k}_T=\frac{1}{2}{\bf
p}_T^{\rm miss}$, the solid line is the distribution over the full
event set for $\tilde{\bf k}_T={\bf k}_T^{\rm maos}$, and finally
the shaded region represents the distribution over the 10\% subset
near the $M_{T2}$ endpoint for $\tilde{\bf k}_T={\bf k}_T^{\rm
maos}$. Our result clearly shows that the MAOS momenta provide a
reasonable approximation to the true WIMP momenta even for the full
event set, and the approximation can be systematically improved by
selecting an event subset near the $M_{T2}$ endpoint.

\vskip -0.5cm
\begin{figure}[ht!]
\begin{center}
\includegraphics[width=15pc]{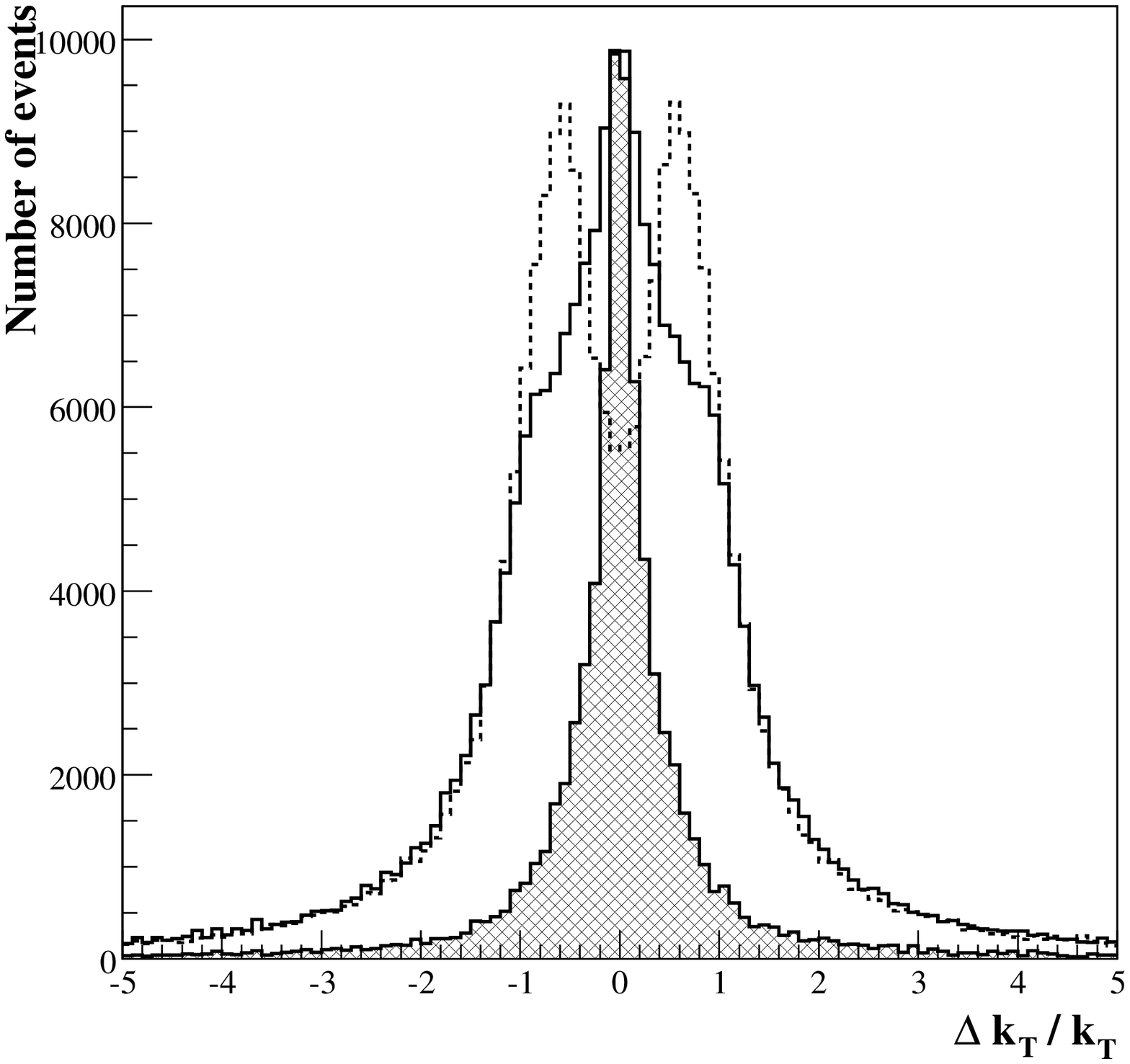}
\end{center}
\vskip -1.3cm \caption{Distribution of $\Delta{\bf k}_T/{\bf
k}_T^{\rm true}$.} \label{susy_dalitz}
\end{figure}

The MAOS momentum constructed as above can be useful for various
purposes. Here we discuss some applications to  spin measurement
\cite{maos}. As the first application,  let us consider the
symmetric 3-body decays of gluino pair in SUSY model: \bea
\tilde{g}+\tilde{g}\rightarrow q(p_1)\bar{q}(p_2)\chi(k)
+q(q_1)\bar{q}(q_2)\chi(l),\nonumber\eea and also the similar decays
of KK gluon pair in UED-like model \cite{csaki}. An observable which
clearly shows the spin effect, so distinguishes the gluino decay
from the KK-gluon decay, is the invariant mass distribution
$d\Gamma/dsdt_{\rm maos}$ for \bea
%\frac{d\Gamma}{dsdt_{\rm maos}}:\,
s=(p_1+p_2)^2,\quad t_{\rm maos}=(p_i+k^{\rm maos})^2, \nonumber
\eea where $p_i$ can be any of $p_1$ and  $p_2$. To be specific, we
choose the focus (SPS2) point of mSUGRA scenario,
%with $m_0
%= 1450$ GeV, $m_{1/2} = 300$ GeV, $A_0=0$, $\tan \beta = 10$, $\mu
%>0$,
%which gives  the following weak scale masses: $m^{\rm
%true}_{\tilde{g}} = 779$ GeV, $m^{\rm true}_\chi = 122$ GeV, and
%$m^{\rm true}_{\tilde q} \simeq 1.5$ TeV. We also consider
and its UED equivalent in which the gluino is replaced with the
first KK gluon, the Bino LSP  with the first KK $U(1)_Y$ boson, and
squarks with the first KK quarks. Using {\tt MadGraph/MadEvent}, we
have generated the events at parton-level for both SUSY and UED
cases, and constructed the MAOS momenta of each event. The resulting
distributions of $s$ and $t_{\rm maos}$ for the gluino 3-body decay
and the KK-gluon 3-body decay at parton-level are depicted in Fig.~5
and Fig.~6, respectively, which clearly reveal the difference
arising from spin effects. This difference survives, at least
qualitatively, even after various errors are taken into account,
including the combinatoric errors and detector effects \cite{maos}.

%\vskip -0.5cm
\begin{figure}[ht!]
\begin{center}
\includegraphics[width=15pc]{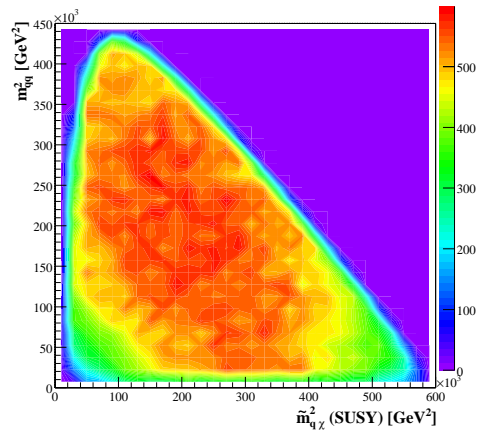}
\end{center}
\vskip -1.3cm \caption{Dalitz Distribution $d\Gamma/dsdt_{\rm maos}$
of the gluino 3-body decay: $s=m_{qq}^2$, $t_{\rm
maos}=\tilde{m}^2_{q\chi}$.} \label{susy_dalitz}
\end{figure}

\begin{figure}[ht!]
\begin{center}
\includegraphics[width=15pc]{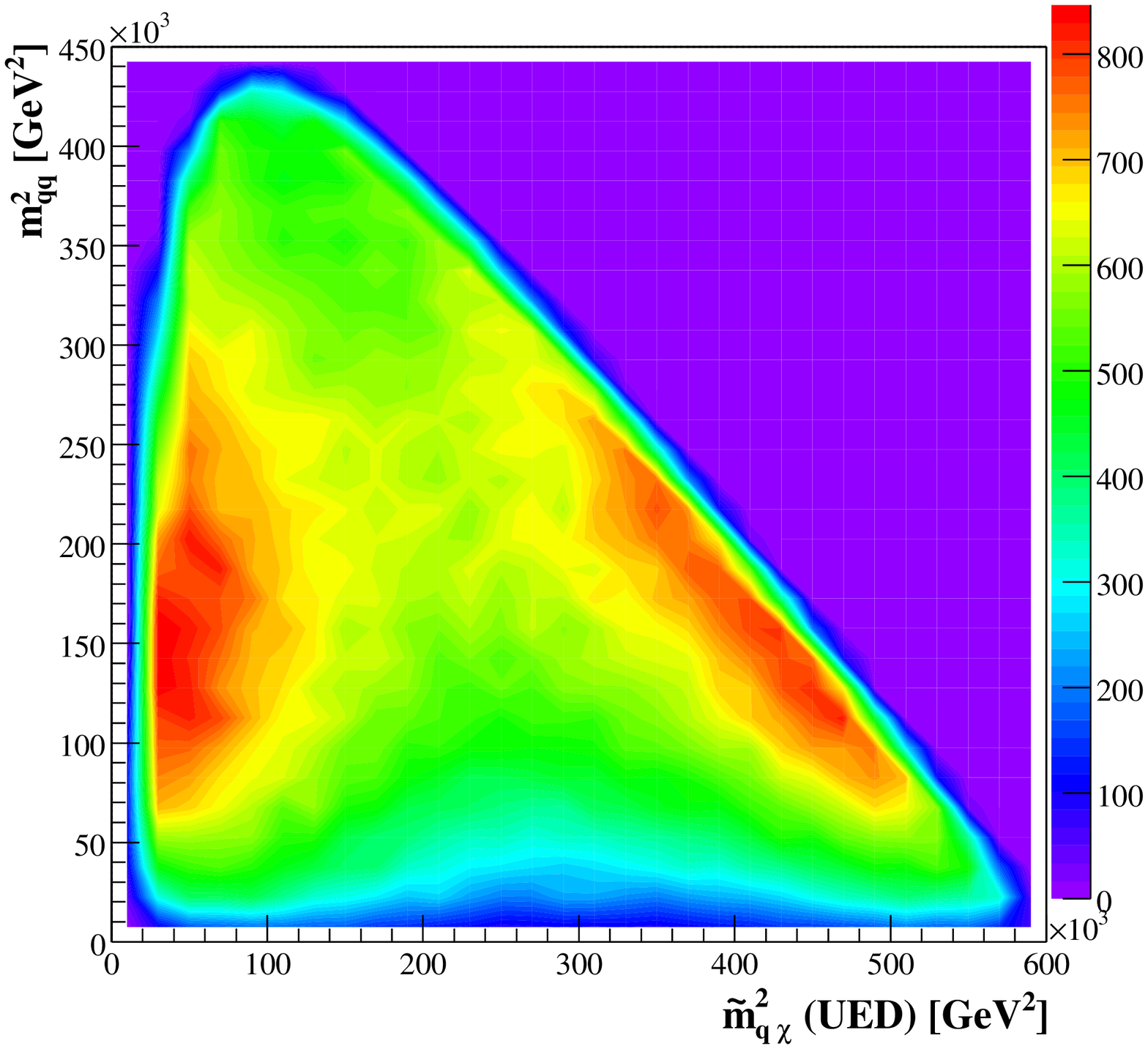}
\end{center}
\vskip -1.3cm \caption{Dalitz Distribution  $d\Gamma/dsdt_{\rm
maos}$ of the KK-gluon 3-body decay: $s=m_{qq}^2$, $t_{\rm
maos}=\tilde{m}^2_{q\chi}$.} \label{ued_dalitz}
\end{figure}

%{ued_dalitz_maos.eps,width=6cm,height=6cm,angle=0} \caption{SUSY and
%UED Dalitz plots of $m_{qq}^2$ and $\tilde{m}_{q\chi}^2$ for
%$m_{\chi,Y}=m^{\rm true}_{\chi,Y}$ and ${\cal L}=300\, {\rm
%fb}^{-1}$.} \label{fig:dalitz}
%

Our next application of the MAOS momentum is the spin determination
of the slepton $\tilde{\ell}$ or of the KK lepton $\ell_{(1)}$ with
the Drell-Yan pair production \cite{barr}:\bea q\bar{q}\rightarrow
Z^0/\gamma\rightarrow Y+\bar{Y} \rightarrow \ell(p)\chi(k)
+\bar\ell(q)\chi(l),\nonumber\eea where $Y=\tilde{\ell}$ or
$\ell_{(1)}$. As slepton is a scalar particle, the angular
distribution is proportional to $1-\cos^2\theta^*,$ where $\theta^*$
is the production angle with respect to the proton beam direction.
On the other hand, the corresponding Drell-Yan production of KK
leptons shows the characteristic distribution of spin-half
particles, which is proportional to
$1+\cos^2\theta^*(E_{\ell}^2-m_{\ell}^2)/(E_{\ell}^2+m_{\ell}^2).$

Once $k^{\rm maos}$ and $l^{\rm maos}$ are obtained, we can probe
the angular distribution of the mother particle MAOS momenta,
$p+k^{\rm maos}$ and $q+l^{\rm maos}$. To see this, we have
generated the events for the SPS1a SUSY point
%($m_{\tilde l_R}^{\rm true}=143$ GeV, $m_\chi^{\rm true}=96$ GeV)
and its UED equivalent with the integrated luminosity $\int {\cal
L}dt=300\, {\rm fb}^{-1}$, and examined the angular distribution of
the mother particle MAOS momentum in the center of mass frame.
Fig.~7 shows the distributions obtained by including all four
different combinations of MAOS momenta, i.e. $(k^{\rm
maos}_\mu(\alpha),l^{\rm maos}_\mu(\beta))$ with $\alpha,\beta=\pm$,
for each event. Here we have employed an event selection adopting
only the top 30\% of events near the $M_{T2}$ endpoint. To see the
efficiency of the MAOS momentum method, we provide also the angular
distributions obtained from the true WIMP momenta. The result shows
that the MAOS angular distribution  reproduces excellently the true
production angular distribution, with which one can distinguish the
slepton production from the KK-lepton production.

\begin{figure}[ht!]
\begin{center}
\includegraphics[width=15pc]{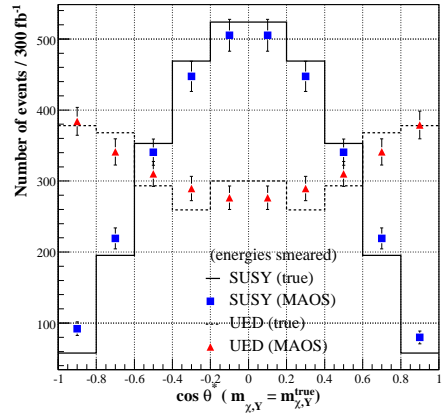}
\end{center}
\vskip -1.3cm \caption{MAOS production angle distribution of slepton
and KK-lepton pairs.} \label{ued_dalitz}
\end{figure}

\section{HIGGS MASS MEASUREMENT WITH MAOS MOMENTUM}

For the SM Higgs boson in the mass range $135 \,\,{\rm GeV}\leq
m_H\leq 180 \,\, {\rm GeV}$, the Higgs decay \bea H \to W+W \to
\ell(p)\nu(k)+ \bar\ell(q) \nu(l),\eea with $\ell =e\,,\mu$ may
provide the best search channel.
% Even for a heavier Higgs boson with
%$m_H \geq 2 M_Z$, this channel gets benefit from a larger branching
%ratio compared to the decay
% $H \to ZZ \to 2\ell 2\bar\ell$.
Certainly this is a type of process that the MAOS momentum can be
employed to measure the unknown Higgs mass, or to discover/exclude
the Higgs boson in certain mass range. However, in case that
$m_H<2M_W$, one of the W-bosons should be in off-shell, and then the
first MAOS scheme using the solution of (\ref{tmaos}),
(\ref{missingpt}) and (\ref{maos1}) does not provide an acceptable
approximation to the true neutrino momenta. We find that the second
scheme using the solution of (\ref{tmaos}), (\ref{missingpt}) and
(\ref{maos2}) works well regardless of whether $m_H>2M_W$ or not
\cite{maoshiggs}, although it might not be as good as the first
scheme when $m_H>2M_W$. We thus choose the second scheme to obtain
the neutrino MAOS momenta in the dileptonic decays of W-boson pair,
and apply it to the Higgs mass measurement.

Once the MAOS momenta of neutrinos are obtained, one can construct
the MAOS Higgs mass:
\begin{equation}
\left(m_H^{\rm maos}\right)^2 \equiv (p + {k}_{\rm maos} + q +
{l}_{\rm maos})^2\,.
\end{equation}
The discussion made in the previous section suggests that $m_H^{\rm
maos}$ has a peak at the true Higgs boson mass, which becomes
narrower under an event selection choosing only the events near the
$M_{T2}$ endpoint. Interestingly, such an $M_{T2}$ cut is useful in
another sense as it enhances the signal to background ratio for the
process in consideration.

 To examine
the experimental performance of the reconstruction of $m_H^{\rm
maos}$ at the LHC, we have generated the Monte Carlo event samples
of the SM Higgs boson signal and the two main backgrounds, using
{\tt PYTHIA6.4} while assuming the integrated luminosity of 10
fb$^{-1}$. The generated events have been further processed through
the fast detector simulation program {\tt PGS4} to incorporate the
detector effects. For the signal, we consider the Higgs boson
production via the gluon fusion: $gg\rightarrow H$. The dominant
background comes from the continuum $q\bar{q}\,,\,gg \to WW \to l\nu
l^\prime\nu^\prime$ process, and we include also the $t\bar{t}$
background in which the two top quarks decay into a pair of $W$
bosons and two $b$ jets.

It is well known that the background can be significantly reduced by
exploiting the helicity correlation between the charged lepton and
its mother $W$ boson. Introducing the transverse opening angle
between  two charged leptons, $\Delta\Phi_{ll}$, the Higgs signal
tends to have a smaller $\Delta\Phi_{ll}$ than the background, which
is essentially due to the fact that the Higgs boson is a spin zero
particle.
% behaves differently to prefer
%$\Delta\Phi_{ll}=\pi$.
%
In fact, there is a correlation between $\Delta\Phi_{\ell\ell}$ and
$M_{T2}$
%which is
%valid for dileptons from generic $W$ pair system with small
%transverse momentum,
in such a way that the dileptons from the Higgs decay are likely to
have larger $M_{T2}$ than the background \cite{maoshiggs}. As a
result, selecting the events with large value of $M_{T2}$ also
enhances the signal to background ratio. In our case, this $M_{T2}$
cut is particularly useful since it enhances also the accuracy of
the MAOS reconstruction of the neutrino momenta as discussed in the
previous section.

We have imposed the usual selection cuts on the Higgs signal and the
backgrounds, and then incorporate additional cuts: $\Delta\Phi_{ll}
< \Delta\Phi_{ll}^{\rm cut}$ and $M_{T2} > M_{T2}^{\rm cut},$ where
$\Delta \Phi_{ll}^{\rm cut}$ and $M_{T2}^{\rm cut}$ are chosen to
optimize the Higgs mass measurement.  In Figs.~8 and 9, we show the
resulting distribution of ${m}_H^{\rm maos}$ for the input mass
$m_H=150$ GeV and $m_H=180$ GeV, respectively. Each distribution has
a clear peak over the background at the true Higgs mass, which
suggests that one might be able to determine  the Higgs boson mass
accurately with a template fitting to the MAOS Higgs mass
distribution. A detailed likelihood fit analysis has been made in
\cite{maoshiggs}, and the results indicate
 that the MAOS Higgs mass distribution indeed
gives a better determination of the Higgs boson mass than other
kinematic variables \cite{bgl}.

\begin{figure}[ht!]
\begin{center}
\includegraphics[width=15pc]{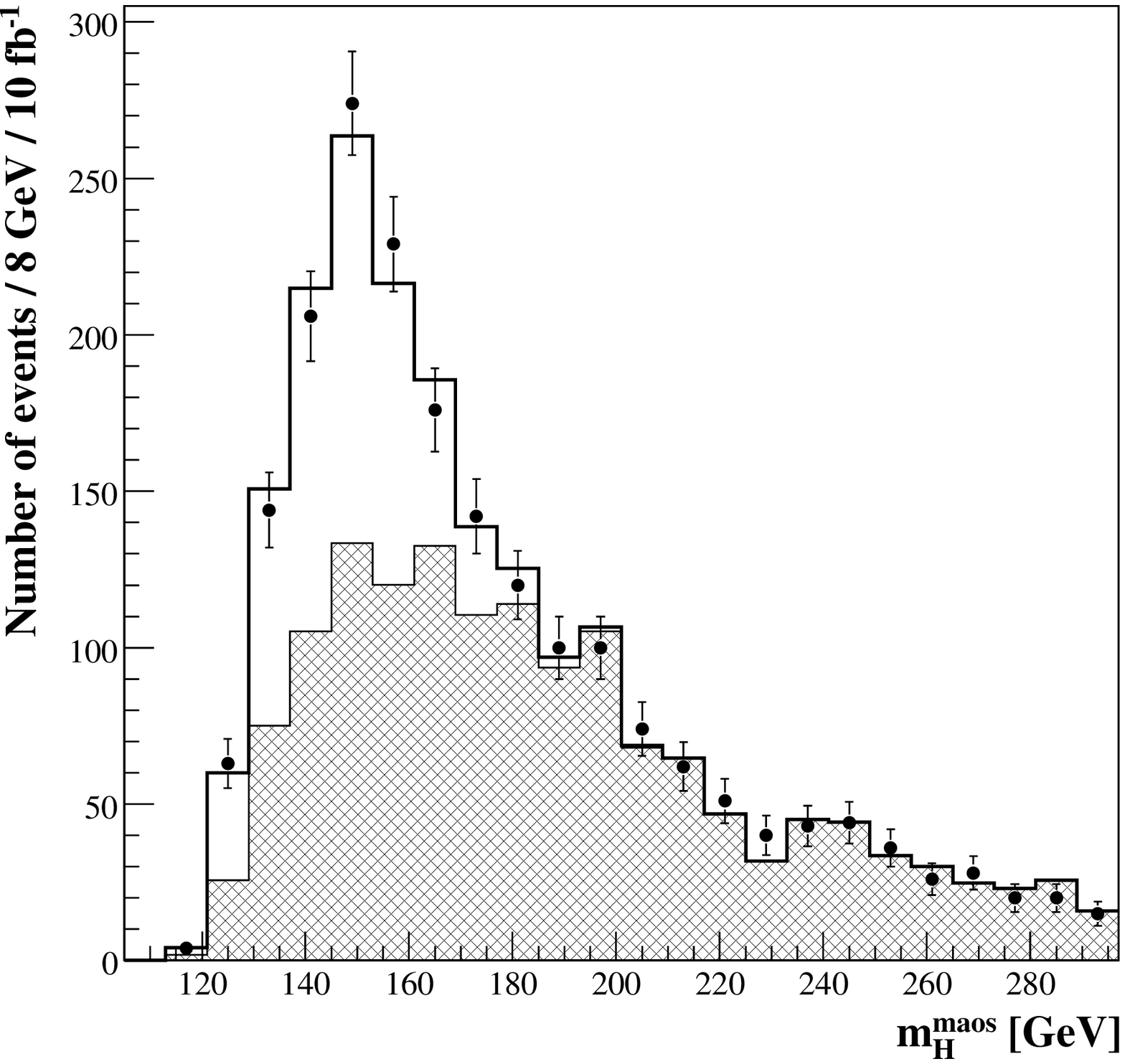}
\end{center}
\vskip -1.3cm \caption{MAOS Higgs mass distribution for $m_H=150$
GeV. Shaded region represents the backgrounds.} \label{mh150}
\end{figure}

\begin{figure}[ht!]
\begin{center}
\includegraphics[width=15pc]{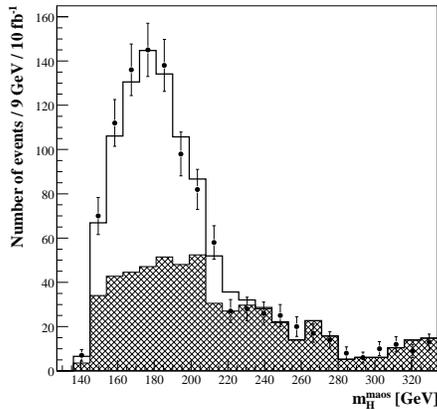}
\end{center}
\vskip -1.3cm \caption{MAOS Higgs mass distribution for $m_H=180$
GeV.} \label{mh180}
\end{figure}

\section{CONCLUSION}

In this talk, we have discussed the recently proposed $M_{T2}$-kink
method to determine the unknown masses in new physics events with
missing energy. Unlike the other methods such as the endpoint method
and the mass relation method, the $M_{T2}$-kink method does not
require a long chain of decays, so can be applied in principle to a
wide class of new physics processes producing a pair of invisible
WIMPs in the final state. We also introduced a new kinematic
variable, the $M_{T2}$-Assisted-On-Shell (MAOS) momentum,  providing
a systematic approximation to the invisible particle momenta in such
processes. We then discussed some applications of the MAOS momentum,
which would determine the gluino or slepton spin, and also the Higgs
boson mass in the process $H\rightarrow WW\rightarrow
\ell\nu\bar{\ell}\nu$. Still much works remain to be done to see in
which other cases the $M_{T2}$-kink method or the MAOS momentum can
be successfully employed to measure the unknown mass or spin with
real collider data.


\begin{thebibliography}{9}
\bibitem{susy}
H. P. Nilles, {Phys. Rept.} {110} (1984) 1; H. E. Haber and G. L.
Kane, {Phys. Rept.} {117} (1985) 75.

\bibitem{littlehiggs}
H. C. Cheng and I. Low, {JHEP} {0309} (2003) 051 [{
hep-ph/0308199}].

\bibitem{ued}
T. Appelquist, H. C. Cheng and B. A. Dobrescu, {Phys. Rev.} {D64}
(2001) 035002 [{hep-ph/0012100}].

\bibitem{mt2kink1}
W. S. Cho, K. Choi, Y. G. Kim and C. B. Park, {Phys. Rev. Lett.}
{100} (2008) 171801 [{arXiv:0709.0288}]; W. S. Cho, K. Choi, Y. G.
Kim and C. B. Park, {JHEP} {0802} (2008) 035 [{arXiv:0711.4526}].
\bibitem{mt2kink2}
 B. Gripaios,
  JHEP {0802}, 053 (2008)
  [{arXiv:0709.2740}];
A. J. Barr, B. Gripaios and C. G. Lester, {JHEP} {0802} (2008) 014
[{arXiv:0711.4008}].
\bibitem{maos}
  W.~S.~Cho, K.~Choi, Y.~G.~Kim and C.~B.~Park,
  %``M_T2-assisted on-shell reconstruction of missing momenta and its
  %application to spin measurement at the LHC,''
  Phys. Rev.  {D79} (2009) 031701
  [arXiv:0810.4853].
\bibitem{maoshiggs}
K.~Choi, S.~Choi, J.~S.~Lee, and C.~B.~Park, arXiv:0908.0079
[hep-ph].

\bibitem{endpoint} I. Hinchliffe, F. E. Paige, M. D. Shapiro, J.
Soderqvist, and W. Yao, Phys. Rev. D55 (1997) 5520 [hep-ph/9610544];
%H. Bachacou, I. Hinchliffe, and F. E. Paige, Phys. Rev. D62 (2000)
%015009 [hep-ph/9907518];
For a recent study, see K. T. Matchev, F.
Moortgat, L. Pape, and M. Park, arXiv:0906.2417 [hep-ph].



\bibitem{mass_relation}
M. M. Nojiri, G. Polesello, and D. R. Tovey, hep-ph/0312317; H. C.
Cheng, D. Engelhardt, J. F. Guinion, Z. Han, and B. McElrath, Phys.
Rev. Lett. 100 (2008) 252001 [arXiv:0802.4290].

\bibitem{gaugino_code}
K. Choi and H. P. Nilles, JHEP {0704} (2007) 006 [hep-ph/0702146].



\bibitem{lester}
C. G. Lester and D. J. Summers, {Phys. Lett.} {B463} (1999) 99 [{
hep-ph/9906349}]; A. J. Barr, C. G. Lester and P. Stephens, {J.
Phys.} {G29} (2003) 2343 [{hep-ph/0304226}].



\bibitem{mt2application}
  %G.~G.~Ross, and M.~Serna,
  %``Mass determination of new states at hadron colliders,''
  %Phys.\ Lett.\ {\bf B665} (2008) 212 [arXiv:0712.0943 [hep-ph]];
  M.~M.~Nojiri, Y.~Shimizu, S.~Okada and K.~Kawagoe,
  %``Inclusive transverse mass analysis for squark and gluino mass
  %determination,''
  JHEP {0806} (2008) 035 [arXiv:0802.2412];
  W.~S.~Cho, K.~Choi, Y.~G.~Kim and C.~B.~Park,
  %``Measuring the top quark mass with $m_{T2}$ at the LHC,''
  Phys.\ Rev. \ {D78} (2008) 034019 [arXiv:0804.2185];
  A.~J.~Barr, G.~G.~Ross and M.~Serna,
  %``The Precision Determination of Invisible-Particle Masses at the
  %LHC,''
  Phys.\ Rev.\ {D78} (2008) 056006 [arXiv:0806.3224];
  %M.~M.~Nojiri, K.~Sakurai, Y.~Shimizu and M.~Takeuchi,
  %``Handling jets $+$ missing $E_T$ channel using inclusive
  %$m_{T2$,''
  %JHEP {0810} (2008) 100 [arXiv:0808.1094];
  H.-C.~Cheng and Z.~Han,
  %``Minimal Kinematic Constraints and $m_{T2}$,''
  JHEP {0812} (2008) 063 [arXiv:0810.5178];
  M.~Burns, K.~Kong, K.~T.~Matchev and M.~Park,
  %``Using Subsystem $M_{T2}$ for Complete Mass Determination in Decay
  %Chains with Missing Energy at Hadron Colliders,''
  JHEP {0903} (2009) 143 [arXiv:0810.5576].
%
\bibitem{nojiri}
J.~Alwall, K.~Hiramastsu, M.~.~Nojiri, and Y.~Shimizu,
arXiv:0905.1201 [hep-ph].

\bibitem{csaki}
C. Csaki, J. Heinonen, and M. Perelstein, JHEP {0710} (2007) 107
[arXiv:0707.0014].



\bibitem{barr}

A. J. Barr, JHEP {0602} (2006) 042 [hep-ph/0511115].

\bibitem{bgl}

A. J. Barr, B. Gripios and C. G. Lester, JHEP 0907 (2009) 072
[arXiv:0902.4864].


\end{thebibliography}
\end{document}